\documentclass[prd,aps,preprintnumbers,showkeys,showpacs,amsmath,amssymb,superscriptaddress,nofootinbib]{revtex4}
\usepackage{amsmath,amssymb}
\usepackage{eucal}
\usepackage{color}
\usepackage[dviwindo]{graphicx}
\usepackage{bmpsize}
\usepackage{dcolumn}
\topmargin=.cm
\newcommand{\bs}{\begin{subequations}}
\newcommand{\es}{\end{subequations}}
\newcommand{\ben}{\begin{eqnarray}}
\newcommand{\een}{\end{eqnarray}}
\newcommand{\la}{\label}

\begin{document}

\title{New Results for Quasi Normal Modes of Gravitational Waves }

\author{Plamen P Fiziev}

\affiliation{Sofia University Foundation for Theoretical and Computational Physics and Astrophysics, Boulevard
5 James Bourchier, Sofia 1164, Bulgaria}
\affiliation{BLTF, JINR, Dubna, 141980 Moscow Region, Russia}
\date{\today}

\begin{abstract}
We briefly consider the data of collaboration LIGO/VIRGO for gravitational waves (GW)
and the recent observations of Event Horizon Telescope (EHT),
and we discuss difficulties for finding the right theory of gravity
and the nature of the observed Extremely Compact Objects (ECOs).
The only undisputable way to establish existence of event horizon of ECOs,
is the extraction of Quasi Normal Modes (QNMs) from the ringing phase of the sources of GW.
We present our method for calculation of QNMs of GW in
the Schwarzschild metric with different boundary conditions.
It is based on exact solutions of the Regge-Wheeler and Zerilli equations in terms of
the confluent Heun functions. We present also new  numerical results of high precision ($ \geq 16$ digits) for QNM frequencies.
We indicate the difficulties for comparison of theoretical results for QNMs with the observations.



\end{abstract}

\maketitle

1. {\bf Introduction}
\vskip 0.1truecm

LIGO detection of GW is
the most important finding in gravity
after the Sir Isaac Newton's discovery of gravitational field. Indeed,
Sir Isaac Newton discovered gravitational field attached to bodies.
LIGO discovered gravitational field detached from bodies and freely spreading in space.
The last being a qualitatively different novel gravitational phenomenon.

Without any doubt, with the LIGO/VIRGO observations and analysis of 11 well-established GW events, included in the First catalog of GW events,
and more than 35 additional candidates of such events from the third scientific run,
(see the references at the WEB address \cite{LOGOOPEN}), a new era in fundamental physics  started \cite{Barack2018}.
Finally, the  gravitational astronomy opened a novel window to the Universe and started to give us unreachable until now knowledge about the Nature.

These achievements deserve extraordinary careful analysis of all issues which appear for the first  time  and are accompanied by many uncertainties and unknowns.
Prior ruthless examination of all facts, hypotheses, assumptions, and interpretations we can not be sure what we really see in this newly opened window.

We certainly may refer to the already existing models and theoretical achievement
like well tested in other physical domains General Relativity (GR)
or to the variety of strictly speaking still hypothetical Black Hole (BH) models.

However, the most important in the new situation is the newly appeared opportunity to examine experimentally the theoretical
assumptions, to look for new developments and unexpected physical phenomena.

The used methods for processing the LIGO data were good enough to discover GW without any doubt,
but fail to  recover the most important details needed to establish definitely the right theory
of the observed GW and the physical nature of their sources.

For example, these methods were too crude to recover QNMs as a fingerprints of BH,
see for example \cite{Starobinskij1973,ChandraDetweiler1975,Chandrasekhar1983,Kokkotas1999, Nagar05,Berti06,Fiziev2006,Berti09,Fiziev2011,Barack2018,Testing2019}
and huge amount of references therein.

The used methods also turned out to be not enough to establish existence
or absence of echoes as a fingerprints of ECOs\footnote{In the last few years the abbreviation "ECOs" is used for
Exotic Compact Objects which are not BH. We use here this abbreviation in wider sense, including
BH in the class of ECOs $\equiv$ Extremely Compact Objects.}
which are not BH \cite{Cardoso16,Mark2017,Cardoso2017,Conklin18,Wang2018,Raposo18,Cardoso2019}.

As a result, we still can not refute firmly many of the alternative theoretical possibilities for explanation of the observed by LIGO/VIRGO collaboration GW events,
see the talk by Alan Weinstein at The First LIGO Open Data Workshop 2018 \cite{LOGOOPEN} and \cite{Barack2018}.

The most recent observation of the shadow of the ECO
at the center of the elliptic galaxy M87 proves the extremely compact character of the central object \cite{EHT},
but it does not give any indication of the presence or absence of an event horizon of this object, since it is able
to probe only the region well-outside the event horizon.
Thus, we have no rigorous and firm proof that the central object in M87 is a BH.

Being still an elusive physical phenomenon, the QNMs and their enough-precision-extraction from observational data
 are the only indisputable way to confirm or reject rigorously the existence of event horizon of the observed ECOs,
i.e., to confirm or reject ECOs' BH nature which remains thus far hypothetical.

The basic point of the present paper is to put on rigorous basis the treatment of QNMs of BH and other ECOs$\neq$BH,
accompanied by high-precision numerical calculations.

\vskip .3truecm
{\bf 2. Exact Solutions to the Regge-Wheeler and Zerilli Equations for GW.}\la{ExactRWEandZ}
\vskip .1truecm
In the present paper we consider different type of static spherically symmetric ECOs
and standard linear approximation for GW in body's exterior Schwarzschild metric \cite{Chandrasekhar1983}.

Because of total separation of variables, GW can be constructed using proper superposition
of time-radial-waves of form $\Phi(t,r,\omega,l)=e^{i\omega t}R(r,\omega,l)$
with coefficients, which depend on the standard spherical angles $\theta$ and $\phi$,
see for example \cite{Nigel2016} and the references therein.

The radial functions $R(r,\omega,l)$ are solution of the simple ordinary differential equation of second order
\ben
g{\frac {d}{dr}}\left(g{\frac {dR}{dr}}\right)+ \left(\omega^2-V(r)\right)R=0, \quad r\in (1,\infty),
\la{Requation}
\een
where $g=1-1/r = g_{tt}=-1/g_{rr}\in (0,1)$ defines the interesting for us components
of the Schwarzschild metric (in it's standard Hilbert form, see, for example, \cite{Fiziev2019a}) in the observable real domain of the area radius $r\in (1,\infty)$.
We are using the most often adopted normalization $2M=1$
for the Keplerian mass $M$ of the object and standard geometrical units $G=c=r_{G}=1$.
The metric signature is $\{+,-.-,-\}$.

The explicit form of the potentials $V(r)$ in cases of the radial Regge-Wheeler equation (RWE) and the radial Zerilli equation (ZE)
(both of the type of Eq.\eqref{Requation}) can be found, for example, in \cite{Chandrasekhar1983}:
\begin{subequations} \la{potentials:ab}
\ben
V_{RW}(r)&=&g\left( {\tfrac L {r^2}}-{\tfrac 3 {r^3}}\right)= - g{\tfrac {dW}{dr}} +W^2 -|\omega_{as}|^2,\qquad\la{potentials:a}\\
V_{Z}(r)&=&g{\tfrac {L(L-2)^2r^3+3\left( (L-2)r+3/2)\right)^2+9/4}{r^3((L-2)r+3)^2}}= \qquad\\\la{potentials:b}
&=& + g{\tfrac {dW}{dr}} +W^2 -|\omega_{as}|^2,\quad \text{where}\quad L=l(l+1), l=2,3\dots,
\nonumber
\een
\end{subequations}
\ben
 W\!=\!g \left( {\tfrac 1 r}\! -\! {\tfrac {L-2}{(L-2)r +3}} \right)\!+\!|\omega_{as}| \to |\omega_{as}| \,\, \text{for } r \to 1,\infty,\quad
 \text{and} \quad \omega_{as}(l)= {\frac i 6}L(L-2)\hskip 3.truecm \la{psi}
\een
are the pure imaginary frequencies of the algebraically special modes of the Schwarzschild BH.

In terms of the $W$ function, the potentials $V_{RW}$ and $V_{Z}$ are deeply related.
As a result, they have identical spectra for the Schwarzschild BH QNMs.
The analytical proof of this iso-spectral property was sketched briefly in \cite{Chandrasekhar1983}.
Below we confirm the iso-spectral property with high numerical precision.

Note that the iso-spectral property is valid only in spacetime dimension $D=1+3$ and its observational check by QNMs can probe the spacetime dimension.

Under other boundary conditions the spectra of the potentials $V_{RW}$ and $V_{Z}$ are related, but not identical.
Applying our new method, here we study with high precision some of that spectra for the first time.

Note that on real axis of the variable $r$ the two potentials $V_{RW}$ and $V_{Z}$ are quite close to each other even for $l=2$.
For $l> 2$ these potentials are much closer for real values of $r$ and coincide in limit $l\to\infty$ \cite{Chandrasekhar1983}.

For our approach to QNMs, it is very important that as a function of {\em complex} variable $r$ the potentials $V_{RW}$ and $V_{Z}$
are analytic functions with some simple poles in complex plane $\mathbf{C}_r$,
i.e., holomorphic functions (even meromorphic ones) -- in a strict mathematical terminology which we will not use hereafter.

Then, if we need to know all essential properties of the solutions to Eq.\eqref{Requation},
we have to study these solutions in the whole complex plane $\mathbf{C}_r$,
according to analytical theory of ordinary differential equations \cite{Forsyth1902,Golubev50,CL1987}.
This is the essence of our approach to the problem of QNMs.

The natural and ultimate reason to do this is that, due to the specific
boundary conditions, the QNMs, and corresponding solutions of Eq. \eqref{Requation},
are complex, despite the fact that potentials \eqref{potentials:ab} are real.

The general physical reason for this phenomenon is that we are considering
open physical systems and their energy is not conserved, see Sections III -- VIII.

The exact solutions to RWE for waves with different spin $s$ in the Schwarzschild background metric
ware described in detail and used for the first time in \cite{Fiziev2006}, see also \cite{Fiziev2009,Fiziev2011}.

In the case of GW (spin s=2), up to some complex normalization constants $C^{(1)}_{\pm}(r,\omega,l)$,
one obtains local solutions around the event horizon $r=1$ of the radial RWE
\ben
R^{(1)}_{\pm}(r,\omega,l)\!=\!r^3 e^{i\omega r} (r\!-\!1)^{\pm i\omega}\,\text{HC}_{\pm}(r,\omega,l)
\la{RWR1pm}
\een
in terms of the specific for GW confluent Heun's functions
$\text{HC}_{\pm}=\text{HC}_{\pm}(r,\omega,l)= \hskip 3.truecm\\
\text{HeunC}\left(-2i\omega,\pm 2i\omega,4,-2\omega^2,2\omega^2+4-L,1-r\right).$

Using the nontrivial relation between exact solutions of RWE and ZE \cite{Chandrasekhar1983,Fiziev2009},
here we present for the first time the explicit form of the exact solutions to the radial ZE:
\ben
R^{(1)}_{\pm}(r,\omega,l)\!=\!r^3 e^{i\omega r} (r\!-\!1)^{\pm i\omega}\!\left(a_\pm \text{HC}_{\pm}+b_\pm{\tfrac d {dr}}\text{HC}_{\pm}\right)\!,\quad
\la{ZR1pm}
\een
so far also up to corresponding complex normalization constants $C^{(1)}_{\pm}(r,\omega,l)$. We obtain for the coefficients $a_\pm$ and $b_\pm$
\ben
a_{+}=\!L(L\!-\!2)\!+\!{\tfrac {2(l+10)}{r}}\!-\! {\tfrac {24}{r^2}}\!-\!{\tfrac {2(L-2)(L+1)}{(L-2)r+3}}\! + i 6\omega, \qquad
a_{-}=\,a_{+}-i{\tfrac {12\omega}{r}},\qquad b_\pm=-6g.
\la{apm}
\een

In the present  paper  we consider common general properties of the solutions \eqref{RWR1pm} and \eqref{ZR1pm} defined by the common factors
in front of the functions $\text{HC}_{\pm}$ and $\left(a_\pm \text{HC}_{\pm}+b_\pm{\tfrac d {dr}}\text{HC}_{\pm}\right)$, correspondingly,
as well as the common limits of these functions when one approaches boundaries
of the observable interval $r\in (1,\infty)$, i.e. when $r\to 1$ or $|r|\to\infty$ .

These general considerations are the same for the solutions of radial RWE \eqref{RWR1pm} and radial ZE \eqref{ZR1pm}.
Therefore, we use for these solutions the same notations $R^{(1)}_{\pm}(r,\omega,l)$ and $ R^{(\infty)}_{-}(r,\omega,l)$ (See below.).

Taking into account that $\text{HeunC}(...,0)=0$ and $g{\tfrac d {dr}}\text{HC}_{\pm}|_{r=0}=0$,  one sees that for $r>1$ the solution
$\Phi^{(1)}_{+}(t,r,\omega,l)$ describes a GW which enters into horizon, and $\Phi^{(1)}_{-}(t,r,\omega,l)$
describes a GW which goes back outside the horizon to infinity.

Considering complex $r-1=|r-1|e^{i\alpha},\omega=|\omega|e^{i\beta} \in \mathbb{C}$ one obtains in the limit $|r-1|\to 0$
\ben
 { R^{(1)}_{-}(r,\omega,l)} \big/  { R^{(1)}_{+}(r,\omega)} \rightarrow 0, \,\, \text{for} \,\, \beta \in (0,\pi),\la{R1pmLimits:a}\quad
 { R^{(1)}_{+}(r,\omega,l)} \big/ { R^{(1)}_{-}(r,\omega)} \rightarrow 0, \,\, \text{for} \,\, \beta \in (-\pi,0).
 \la{R1pmLimits}
\een

In the first of Eqs. \eqref{R1pmLimits} one has a steepest descent for $\beta={\frac\pi 2}$, and
in the second of Eqs. \eqref{R1pmLimits} -- for $\beta=-{\frac\pi 2}$.

The local solutions around infinity $|r|=\infty$ are given by asymptotic series
\ben
R^{(\infty)}_{\pm}(r,\omega,l)= {\tfrac1 r}e^{\mp i\omega \left({r +\ln(r)}\right)}\sum_{n=0}^\infty a^\pm_n r^{-n}, \quad a^\pm_0=1\quad
\la{R1nfpm}
\een
with different coefficients $a^\pm_n$ for corresponding solutions of RWE and ZE.
These coefficients are not essential for our consideration.

The solution $\Phi^{(\infty)}_{+}(t,r,\omega,l)$ describes a GW that goes from horizon to infinity, and $\Phi^{(\infty)}_{-}(t,r,\omega,l)$
describes a GW that goes back from infinity to horizon.

Considering complex $r=|r|e^{i\alpha},\omega=|\omega|e^{i\beta} \in \mathbb{C}$ one obtains in the limit $|r|\to \infty$
\ben
 { R^{(\infty)}_{-}(r,\omega,l)} \big/ { R^{(\infty)}_{+}(r,\omega,l)} \rightarrow 0, \, \text{for}
 \,\alpha\!+\! \beta \in (0,\pi), \la{RinfpmLimits:a}\quad
  { R^{(\infty)}_{+}(r,\omega,l)} \big/ { R^{(\infty)}_{-}(r,\omega,l)}
  \rightarrow 0, \, \text{for} \, \alpha\!+\!\beta \in (-\pi,0).
  \la{RinfpmLimits}
\een

In the first of the Eqs. \eqref{RinfpmLimits} one has steepest descent for $\alpha+\beta={\frac\pi 2}$, and
in the second of the Eqs. \eqref{RinfpmLimits} -- for $\alpha+\beta=-{\frac\pi 2}$.

Note that relations \eqref{R1pmLimits} and \eqref{RinfpmLimits}
show which solution is large or small in corresponding limits,
depending on the direction in {\em complex} r-plane.

The local solutions \eqref{RWR1pm}, \eqref{ZR1pm}, and corresponding solutions \eqref{R1nfpm} form two different complete sets of solutions to RWE or to ZE.
In the common r-domain of their validity (possibly after proper analytical continuation) one can connect them using (unfortunately so far unknown)
connection coefficients $\Gamma^{(1)\,\pm}_{\pm \,\,\,\infty}$. For example:
\ben
R^{(1)}_{\pm}(r,\omega,l) = \Gamma^{(1)\,+}_{\pm \,\,\,\infty}\, R^{\infty}_{+}(r,\omega,l) + \Gamma^{(1)\,-}_{\pm \,\,\,\infty}\, R^{\infty}_{-}(r,\omega,l).\qquad
\la{connection}
\een

Then, using proper constants $C_{(1)}^{\pm}$, any solution $R$ to the radial RWE can be represented as
\ben \la{R}
R(r,\omega,l) = C_{(1)}^{+} R^{(1)}_{+}(r,\omega,l)  &+&  C_{(1)}^{-} R^{(1)}_{-}(r,\omega,l)=\\
\left( C_{(1)}^{+} \Gamma^{(1)\,+}_{+ \,\,\,\infty}  +  C_{(1)}^{-}\Gamma^{(1)\,+}_{- \,\,\,\infty }\right) R^{\infty}_{+}(r,\omega,l) &+&
\left( C_{(1)}^{+} \Gamma^{(1)\,-}_{+ \,\,\,\infty} \! +\!  C_{(1)}^{-}\Gamma^{(1)\,-}_{- \,\,\,\infty }\right) R^{\infty}_{-}(r,\omega,l).\nonumber
\een

\vskip .3truecm

{\bf 3. The Sommerfeld Boundary Condition}\la{SBCsection}
\vskip .1truecm
The common part of QNMs' definition of all types of ECOs is the Sommerfeld Boundary Condition (SBC).
It allows only presence of going from horizon to infinity GW $\Phi^{(\infty)}_{+}(t,r,\omega,l)$, i.e., in Eq. \eqref{R} we must put
\ben C_{(1)}^{+} \Gamma^{(1)\,-}_{+ \,\,\,\infty} + C_{(1)}^{-}\Gamma^{(1)\,-}_{- \,\,\,\infty }=0
\la{SBC0}
\een

Then the second of the Eqs. \eqref{RinfpmLimits} and Eq.\eqref{R} show that the SBC is equivalent to the requirement
\ben
\lim_{|r| \to \infty} R\left(r=|r|e^{-i\left(arg(\omega)+\pi/2\right)},\omega,l\right) =0.
\la{SBC}
\een
This way we avoid the use of the unknown coefficients $\Gamma^{(1)\,\pm}_{\pm \,\,\,\infty}$,
choosing to take limit $|r|\to \infty$ in a proper optimal direction in the complex plane $\mathbb{C}_r$.

Note that according to the first of the Eqs. \eqref{RinfpmLimits}, if $r\to \infty$ on the real axis,
then the large solution $R^{(\infty)}_{+}(r,\omega,l)$ dominates and we are loosing the small solution
$R^{(\infty)}_{-}(r,\omega,l)$ that is needed for calculation of QNMs.
This is a well-known basic and insurmountable difficulty in the direct attempts for numerical calculations of QNMs
using real $r\in (1,\infty)$  \cite{ChandraDetweiler1975}.

We overcome this problem using the optimal direction in the complex plane $\mathbb{C}_r$
in which $R^{(\infty)}_{-}(r,\omega,l)$ dominates, according to the second of Eq. \eqref{RinfpmLimits}.

The physical meaning of the SBC is clear: It describes a flax of gravitational-wave energy from the source
to infinity without any back flax.

This means, that the Universe with a source of GW in it is considered as an open physical system.
As a result, the energy of the source of GW is not conserved and QNMs are complex numbers.

In addition to SBC, each type of ECO characterizes by some specific second boundary condition
in the observable domain, i.e.,  at finite $r_0\geq 1$ which defines its QNM spectrum together with SBC  Eq.\eqref{SBC}.

\vskip .3truecm
{\bf 4. GW-QNM spectrum of Schwarzschild BH}\la{BHQNM}
\vskip .1truecm
The numerical investigation of the GW-QNMs of Schwarzschild BH (SBH) making use of the exact solutions
Eqs.\eqref{RWR1pm} and \eqref{R1nfpm} was started in \cite{Fiziev2006,Fiziev2011}.

In this case, one has to consider a specific class of solutions which describe simultaneously ingoing into the horizon GW
$\Phi^{(1)}_{+}(t,r,\omega,l)$ - BH Boundary Condition (BHBC), together with SBC. The BHBC gives
$C_{(1)}^{-}=0$ in Eq.\eqref{R}.

Thus, from Eq.\eqref{SBC} one obtains the BHBC spectral condition for Swarzschild GW-QNMs
\ben
 \text{HC}_{+}\left( -i |r_\infty| {\tfrac {|\omega|}\omega, \omega}, l  \right)=0.\qquad
\la{BHSQNM}
\een

From a physical point of view, BHBC \eqref{BHSQNM} means that we are considering the observable domain
as an open physical system also at the horizon $r=1$.

As a result, the observable domain of the SBH is a doubly opened physical system, i.e.,
opened at the two ends of the interval $r\in (1,\infty)$.
This physical property is independent of the choice of coordinates and defines the very SBH.

Solutions of the linear perturbations of such physical system
exist only for an unique infinite discrete series of frequencies
$\omega_{n, l}$, $n=0,1,2\dots$, $l=2,3,4,\dots$

\begin{figure}[!ht]
\centering
\begin{minipage}{15.truecm}
\vskip .truecm
\hskip -0.8truecm
\includegraphics[width=0.4\columnwidth]{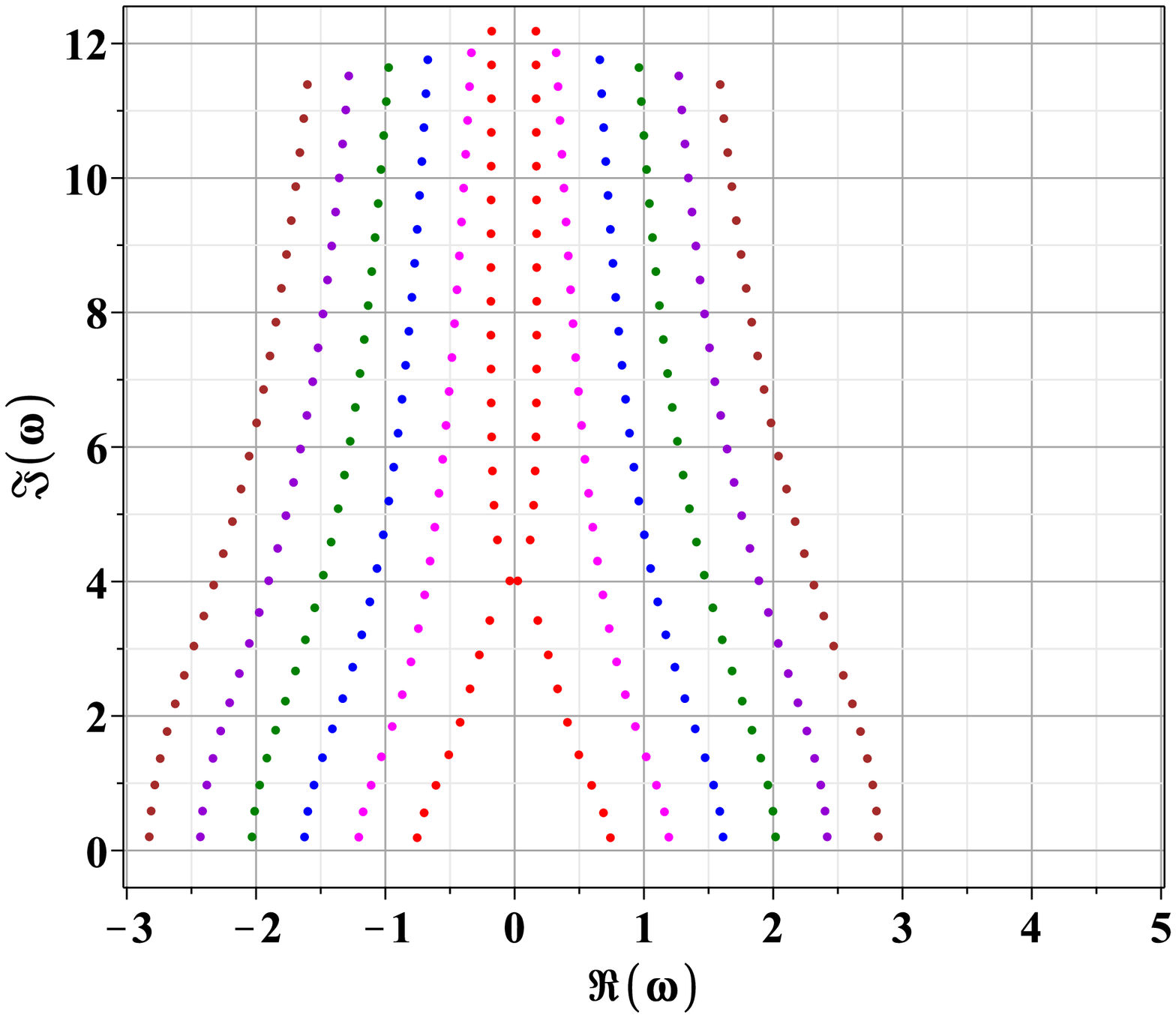}
\begin{minipage}{10.truecm}
\vskip -8.95truecm
\hskip 2.95truecm
\includegraphics[width=0.20\columnwidth]{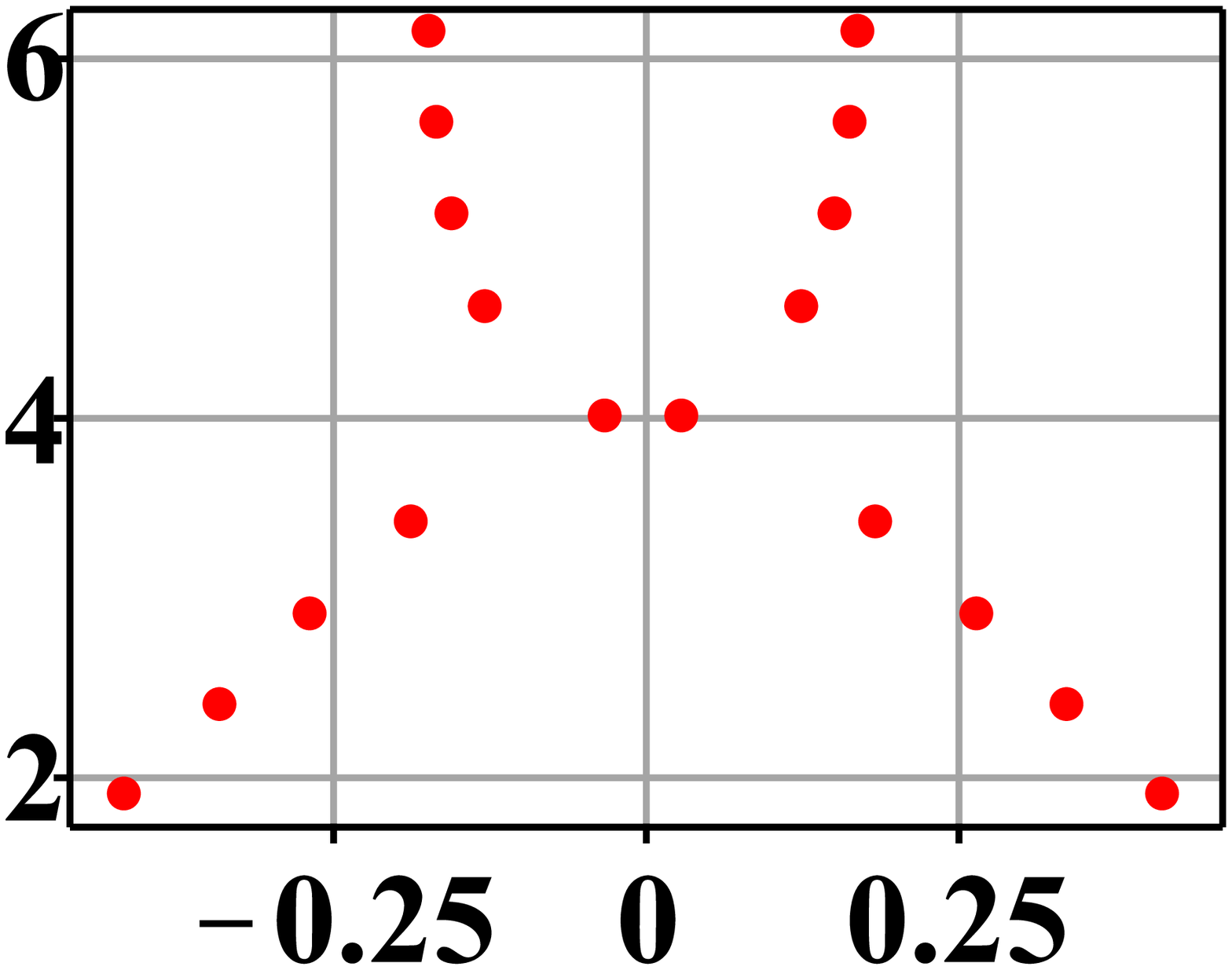}
\end{minipage}
\vskip -.5truecm
\caption{\small The GW BHSBC QNMs, $n=0,\dots24$, $l=2,3,4,5,6,7$ (red, magenta, blue, green, violet, orange)}
\label{Fig1}
\end{minipage}
\end{figure}

The initial 150 in number QNM frequencies $\omega_{n, l}$ for $n=0,\dots,24, l=2,\dots, 7$ , obtained numerically from Eqs.\eqref{SBC} and \eqref{BHSQNM},
are shown in  Fig.\ref{Fig1}. All these frequencies were obtained for the first time with precision $\geq$ 16-digits.
Our 28-digits-result for the basic mode is
\ben
\omega_{0,2}=\pm \, 0.7473433688360836715869840059 + 0 .1779246313778713965609218543\, i.
\la{om02}
\een

In the last months, there appeared two papers \cite{Hatsuda2019,Matyjasek2019} with high precision numerical calculations of QNMs of SBH,
based on high order (up to 200-th and 250-th order) semiclassical approximations,
followed by Borrel, and Borrel-Pad\'e summation of the corresponding asymptotic series.

The  20-digits-result of \cite{Hatsuda2019} and  the 16-digits-result of \cite{Matyjasek2019} for $\omega_{0,2}$
coincide with \eqref{om02} up to their precision. Unfortunately, the higher QNMs, obtined using methods of \cite{Hatsuda2019,Matyjasek2019},
have drastically decreasing precision, in sharp contrast to our results.

The value of the frequency $\omega_{8,2}$ is still under active debate in the literature \cite{Barack2018}. Our 21-digits result
\ben
\omega_{8,2}=\pm \,0.0306490095213001613678 + 3.99682368371788648697 \, i.
\la{om82}
\een
is illustrated in Fig.\ref{Fig1} as an enlarged picture of the SBH-QNM-spectrum around frequency $\omega_{8,2}$.

Here we confirm the original result of \cite{Fiziev2011} about the ninth mode of BHSBC QNMs ($n=8,l=2$) with a higher precision.
According to our results, this mode is {\em not} algebraically special, as often stated in the literature before, as well as after \cite{Fiziev2011}.
This is a result of not-enough-precise calculations and/or incorrect understanding of the role of the algebraically special solutions, which do not satisfy
the boundary conditions for QNMs \cite{Chandra1984}.

Indeed, the numerical value $\omega_{8,2}=0.4615178773933189E\!-\!15 + 3.999999999999607\,i$ with declared precision $0.9154E\!-\!10$:
(see, E. Berti, https://pajes.jh.edu/$\sim$eberti2/ringdown)
can be interpreted as numerically identical to the value $\omega_{as}(2)=0+4\,i$.
However, this does not mean that the BHSBC mode ($n=8,l=2$) is described by algebraically special solution, since it does not obey QNM BC.
The other values of BHSBC frequencies $\omega_{n,l}$ in the Berti tables are in a good agreement with our more accurate ones.

\vskip .3truecm
{\bf 5. GW-QNM of Extremely Compact Objects which are not BH}\la{ECO}
\vskip .1truecm

Consider ECO$\neq$BH with potential
\ben
V(r)=V_{BH}(r)+V_{ref}(r,r_0),\quad \text{where}\quad  V_{BH}=V_{RW},\quad  \text{or}\quad V_{BH}=V_{Z}.
\la{VECO}
\een
In Eq. \eqref{VECO}, a reflecting short-range potential $V_{ref}(r;r_0)$ is added to the Redge-Wheeler or Zerilly potentials.

The place $r_{max}(l)$ of the maxima of the two $V_{BH}$ potentials  decreases monotonically with $l\to\infty$ and has a common limit value
$r_{phs}=1.5$ (the luminosity radius of the photon sphere),
starting from $r_{max}(2)\approx 1.640388$ being the maximum of $V_{RW}$ for $l=2$, and from $r_{max}(2)\approx 1.549395$ being the maximum  of $V_{Z}$ for $l=2$.

\begin{figure}[!ht]
\centering
\begin{minipage}{15.cm}
\vskip -.truecm
\hskip .truecm
\includegraphics[width=.5\columnwidth]{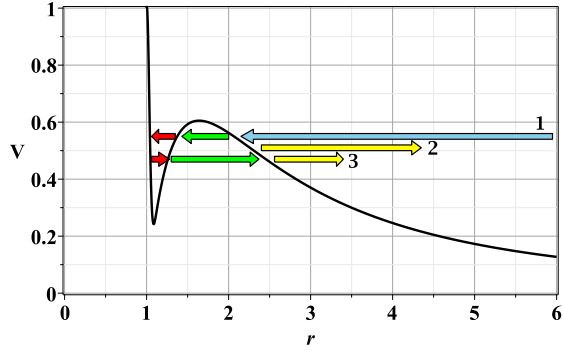}
\vskip .truecm
\caption{\small The potential $V(r)$
and GW: 1 -- Coming from $\infty$ (blue), 2 -- directly reflected by barrier (yellow),
3 -- secondary reflected echo (yellow) after penetration trough the barrier}
\label{Fig2}
\end{minipage}
\end{figure}

We assume the existence of a value $r_0$: $1<r_0< 1.5$ such that $V_{ref}(r;r_0)\equiv 0$ for $r\geq r_0$, see Fig.\ref{Fig2}.

A reflecting potential $V_{ref}(r;r_0)$ \eqref{VECO} with the above properties defines ECO$\neq$BH
and may originate from quantum gravity or from some classical modification of GR, see
\cite{Cardoso16,Cardoso2017,Mark2017,Conklin18,Wang2018} and the references therein.
A specific potential $V_{ref}(r;r_0)$ is defined also by Schwarzschild Massive Point (SMP) solution in GR \cite{Fiziev2019a,Fiziev2019b,Fiziev2019c}.

Due to the Birkhoff theorem, the presence of an additional spherically symmetric source of the potential $V_{ref}(r;r_0)$
does not change the spacetime metric in domain $r\geq r_0$.
It changes only the value of the total mass $M_{tot}= M + \delta M_{ref}$.
The use of standard normalization of the mass $2M_{tot}=1$ makes this change hiden.

Then, in the domain $r \in (r_0,\infty)$ the Eq.\eqref{R} is valid and the reflection coefficient $\mathcal{R}= C_{(1)}^{-}/C_{(1)}^{+}$ is
\ben
\mathcal{R}(\eta)=\mathcal{R}_{DS}\,{\frac{\eta-\eta_{+}}{\eta - \eta_{-}}},\quad \text{where}\quad \mathcal{R}_{DS}=-R^{(1)}_{+}(r_0,\omega,l) /  R^{(1)}_{-}(r_0,\omega,l)
\la{Rref}
\een
is the reflection coefficient for Dirichlet-Sommerfeld-Boundary-Condition (DSBC):
$R(r_0,\omega,l)=0$ \cite{Fiziev2006}.
We use short notation for the logarithmic derivatives
$$\eta_{\pm}=\left({\frac d {dr}}R^{(1)}_{\pm}(r,\omega,l)\right)/R^{(1)}_{\pm}(r,\omega,l)|_{r=r0}\quad \text{and}\quad
\eta=\eta(r_0,\omega,l)=\left({\frac d {dr}}R_<(r,\omega,l)\right)/R_<(r,\omega,l)|_{r=r_0-0}.$$

The function $R_<(r,\omega,l)$ is solution of the problem with potential \eqref{VECO}.
Note that the number $\eta$ is the only trace of the potential \eqref{VECO} in our considerations.

Under proper normalization, the one-parameter family of solutions of the problem in domain $r\in [0,\infty)$ is
\ben
R(r,\omega,l;\eta)= R^{(1)}_{+}(r,\omega,l)+ \mathcal{R}(\eta) R^{(1)}_{-}(r,\omega,l)
\la{Reta}
\een

For $\eta=\eta_{DS}=\infty$ one obtains from formula \eqref{Rref} the Dirichle-Sommerfeld reflection coefficient $\mathcal{R}_{DS}$,
and for $\eta=\eta_{NS}=0$ -- the Neumann-Sommerfeld one $\mathcal{R}_{NS}$, i.e. reflection coefficient for Neumann-Sommerfeld-Boundary-Condition (NSBC):
${\frac d {dr}}R(r_0,\omega,l)=0$
\ben
\mathcal{R}_{NS}=-\left({\frac d {dr}}R^{(1)}_{+}(r_0,\omega,l)\right)\! /\! \left({\frac d {dr}}R^{(1)}_{-}(r_0,\omega,l)\right).\qquad
\la{RNS}
\een
In these two cases we know the explicit form of the reflection coefficient without additional hypothesises.

There exist also one-parameter family of Robin-Sommerfeld Boundary Conditions (RSBC):
$R(r_0,\omega,l)+ \kappa\, {\frac d {dr}}R(r_0,\omega,l) = 0$,
$\kappa \in \mathbb{C}$ being an arbitrary parameter.
The RSBC reflection coefficient
\ben
\mathcal{R}_{RS}=-{\frac {R^{(1)}_{+}(r_0,\omega,l)+ \kappa\, {\frac d {dr}}R^{(1)}_{+}(r_0,\omega,l)}
{R^{(1)}_{-}(r_0,\omega,l)+ \kappa\, {\frac d {dr}}R^{(1)}_{-}(r_0,\omega,l)}}, \quad \text{and}\quad
\eta_{RS}={\frac {\eta_{+}+\mathcal{}{R}_{RS}\,\eta_{-}}{1+\mathcal{}{R}_{RS}}}
\la{kappa}
\een
depend on the parameter $\kappa$, which must be fixed to specify the model.
Note that $\kappa=0$ produces the DSBC, and $\kappa=\infty$ - the NSBC.

Now we can obtain the spectrum of the ECO$\neq$BH-QNMs for any given $\eta$ using the  formula \eqref{SBC}.

\vskip .3truecm
{\bf 6. QNM for GW under DSBC or  NSBC in Schwarzschild metric}\la{DSNSBC}
\vskip .1truecm
The DSBC describes GW with fixed end at the point $r_0$ as a result of the total reflection by some mirror, placed there \cite{Fiziev2006}.
The NSBC describes GW with zero flux and free ends at the point $r_0$ as a result of the reflection by other kind of mirror at the same place.
In these two cases the spectral condition for QNMs of GW is
\ben
\lim_{|r| \to \infty} \big(F(r,r_0,\omega,l)-F(r_0,r,\omega,l)\big)|_{r=-i|r| {\tfrac {|\omega |}\omega}} =0,\quad
\la{DSNSQNM}
\een
where for RWE DSBC $F_{{}_{DS}}(r,r_0,\omega,l)=\left({\frac{r-1}{r_0-1}}\right)^{i\omega}\text{HC}_{+}(r,\omega,l)\text{HC}_{-}(r_0,\omega,l)$.
We skip here the quite complicated explicit expressions of functions $F(r,r_0,\omega,l)$ for other considered by us BC at the point $r_0$.

\begin{figure}[!ht]
\centering
\begin{minipage}{16.cm}
\vskip -.truecm
\hskip -6.truecm
\includegraphics[width=.35\columnwidth]{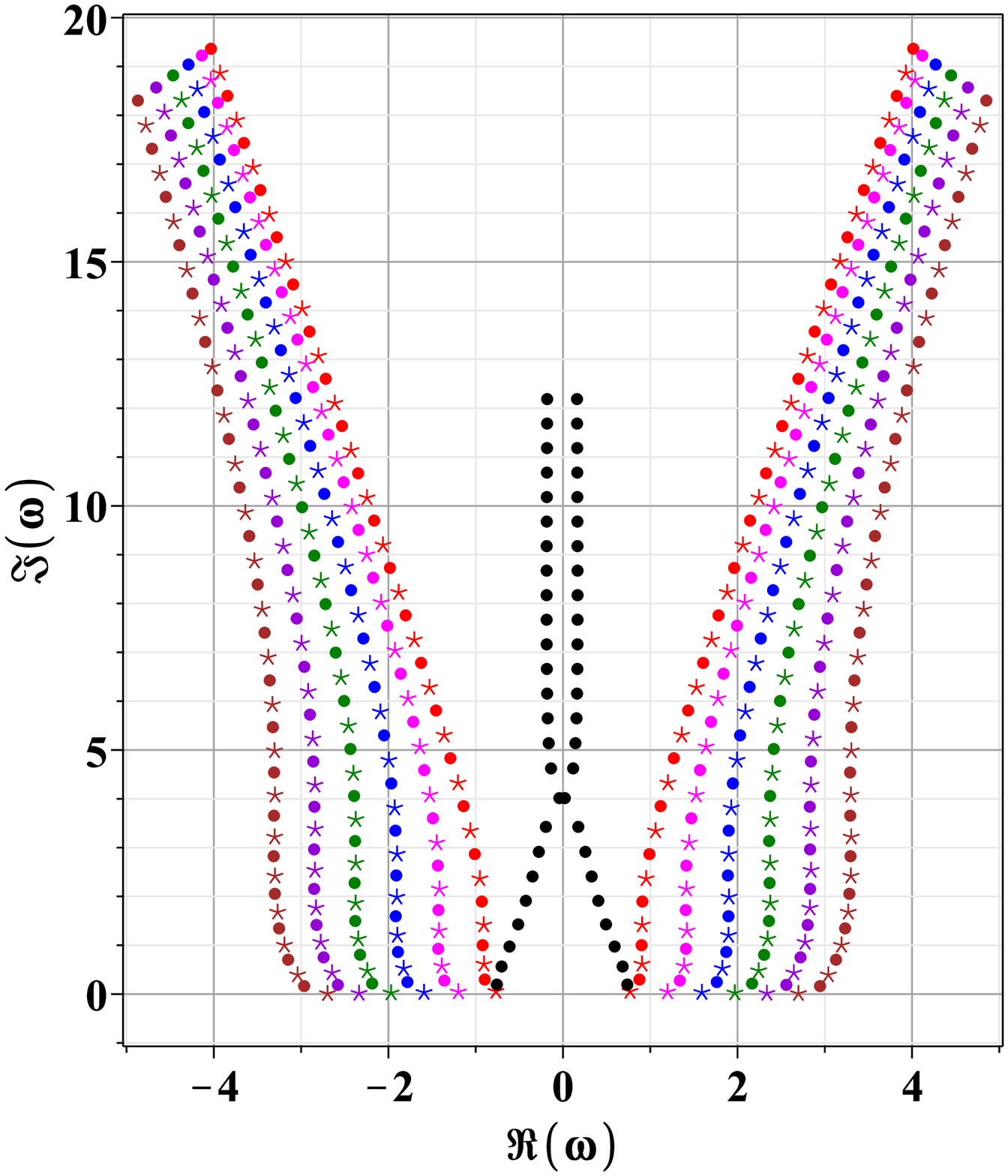}
\end{minipage}
\begin{minipage}{16.cm}
\vskip -6.6truecm
\hskip 6.truecm
\includegraphics[width=.35\columnwidth]{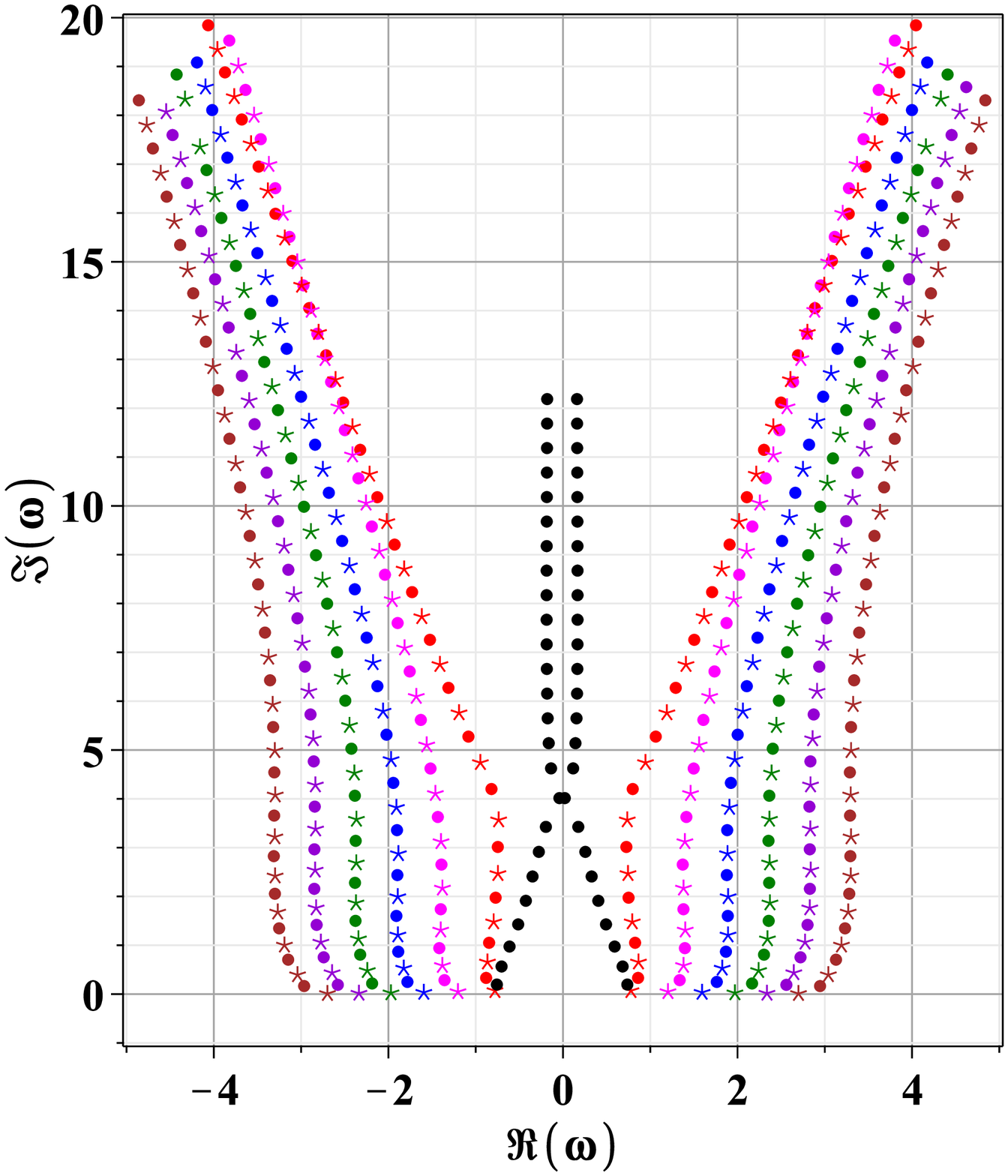}
\end{minipage}
\vskip .truecm
\caption{\small The QNMs of GW  for $r_0=7/6$ , $n=0,\dots, 13$, and $l=2,\dots, 7$ (red, magenta, blue, green,violet);
left: for RWE DSBC; right: for ZE DSBC. The black dots present BH QNMs of GW with $n=0,\dots, 17$ and $l=2$}
\label{Fig3}
\end{figure}

The numerical results for $r_0=7/6$, $|r_\infty| =50+r_0$, $n=0,\dots 13$, and  $l=2,\dots, 7$ are shown in Fig. \ref{Fig3}.

As seen, the QNMs spectra for RWE DSBC(dots), NSBC(stars) and ZE DSBC(dots), NSBC(stars)
are very different from corresponding one for BHSBC, shown in Fig.\ref{Fig1}.
The QNMs spectra for RWE NSBC and ZE NSBC are similar to the RWE DSBC and ZE DSBC spectra, shown in Fig.\ref{Fig3}.
The corresponding QNMs points of RWE NSBC and ZE NSBC are placed at the same lines, approximately at the middle between
the points of  RWE DSBC and ZE DSBC spectra.

Note the obvious change of the behaviour of QNMs under BHSBC, DSBC and NSBC around the value $\Im (\omega)> 3.9968236837$.
The physics behind this phenomenon is still not understood.

The numerical analysis showed also absence of QNMs with negative  imaginary parts $\Im (\omega)$,
thus confirming stability of the BHSBC, DSBC, and NSBC solutions of RWE and ZE.
\begin{figure}[!ht]
\vskip -.truecm
\hskip -.truecm
\includegraphics[width=.5\columnwidth]{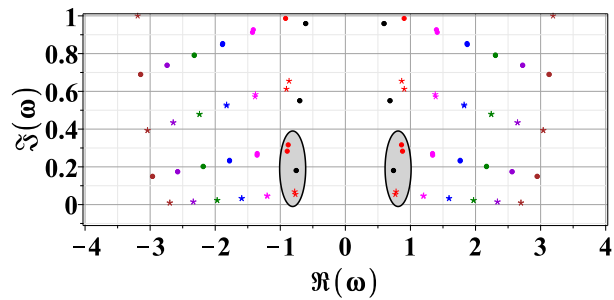}
\vskip .truecm
\caption{\small The basic frequencies of the BHSBC, for RWE and ZE with different BC at the point $r_0$}
\label{Fig4}
\end{figure}
The basic frequencies of the BHSBC, for RWE and ZE coincide, according to isospectral property.
In the case of corresponding DSBC and NSBC we obviously have no isospectral property.
However, the basic QNM frequencies of all these spectra are very close to each other, see Fig. \ref{Fig4}.

Therefore, it seems hoopless to distinguish observationally the basic frequencies of these different spectra in a foreseen future,
even if we succeed to extract firmly QNMs from observational data.

Obviously, we need completely new methods for extracting information about QNMs from the real ECOs, after we detect the radiated by these ECOs GW.



\end{document}